\documentstyle[12pt]{article}

\newcommand{\pl}{\partial_}
\newcommand{\ve}{\varepsilon_}
\newcommand{\ol}{\overline}
\newcommand{\bqq}{\begin{equation} \label}
\newcommand{\eeq}{\end{equation}}
\newcommand{\rr}{{\mbox{\bf R}}}
\newcommand{\rrr}{{\mbox{\small \bf R}}}
\newcommand{\cc}{{\mbox{\bf C}}}
\newcommand{\pp}{{\mbox{\bf P}}}
\newcommand{\tg}{\tilde g}

\newcommand{\tmc}{$TM \otimes_{\mbox{\bf R}}\mbox{\bf C}$}
\newcommand{\dmin}{(\partial_1- \partial_{\overline{1}})}

\newcommand{\appsection}{\addtocounter{section}{1}
           \setcounter{equation}{0}\section*{Appendix \Alph{section}}}

\newtheorem{The}{Theorem}
\newtheorem{Lem}{Lemma}

\begin{document}
\hsize=14.4cm
\textheight=22.5cm

\begin{center}
{\large \bf $H$-PROJECTIVELY EQUIVALENT\\

K\"AHLER MANIFOLDS AND\\
\vskip0.18cm

GRAVITATIONAL INSTANTONS\footnote{
This work was partially supported by Russian Foundation for
Basic Researches (grant No~96-0101031).}
}
\bigskip \bigskip

{\bf Dmitry A. Kalinin}

Department of General\\
Relativity \& Gravitation \\
Kazan State University \\
18 Kremlyovskaya Ul.\\
Kazan 420008 Russia\\
E-mail: dima@tcnti.kazan.ru, dmitry.kalinin@ksu.ru

\end{center}

\small
\section{Introduction}
\normalsize

K\"ahler manifolds were introduced by P.~A.~Shirokov~\cite{shir1} and
E.~K\"ahler \cite{kahl} in the first part of our century. Since that
time they gained applications in a wide variety of fields both in
mathematics and theoretical physics
\cite{al-free,amka2,boraj,flah,kalinin1}. In particular, K\"ahler manifolds 
have been studied as models for finding the gravitational instantons
which are of great importance for construction of quantum
gravity \cite{bess,perry}.

The goal of the present paper is to investigate four dimensional
K\"ahler manifolds admitting
$H$-projective mappings with special attention to
Einstein-K\"ahler manifolds
of this type which can be interpreted as field
configurations of the gravitational instantons.

The notion of $H$-projective mappings was introduced by
T.~Ot\-su\-ki and Y.~Ta\-shi\-ro
\cite{otas} as a generalization of projective
mappings of Riemannian manifolds \cite{am1,sin1}.
At the present moment wide variety of K\"ahler manifolds {\it not
admitting} $H$-projective mappings is known. 
At the same time, some general methods of
finding $H$-projective mappings for given K\"ahler manifold were also
developed \cite{sin1,skm,Yano}.
However, the problem of
finding K\"ahler metrics and connections admitting non-affine
$H$-pro\-jec\-ti\-ve mappings is still unsolved
ever in the case of lower dimensions. Some approaches to its solution
was proposed earlier \cite{amka1,izv2,diss} by the author
in co-laboration with
Prof.~A.~V.~Ami\-nova. 

In the first part of the present paper four-dimensional K\"ahler
manifolds admitting non-affine $H$-projective mappings are studied. 
It is proved that four-dimensional non-Einstein K\"ahler
manifolds admitting $H$-projective mappings are generalized
equidistant manifolds. Moreover, it is proved that
four-dimensional generalized equidistant
K\"ahler manifolds admit $H$-projective mappings in general case. 

The second part of the paper is devoted to investigation of Einstein
generalized equidistant K\"ahler manifolds which can be interpereted 
as field configurations of gravitational instantons.
Explicit expression for the metrics of such
manifolds is found for Ricci-flat case and the case of
Einstein-K\"ahler manifold
($Ric = \kappa g$) with $\kappa\ne 0$. 

\vskip0.2cm

The author is grateful to A.~Ami\-no\-va, K.~Matsumoto and J.~Mi\-ke\v s for
comments, useful discussions and suggestions. My special thanks are
addressed to the referee for valuable remarks and corrections.


\small
\section{Differential geometry of K\"ahler manifolds}
\normalsize

Let me start from reminding some relevant facts on differential
geometry of K\"ahler manifolds \cite{kobn,sin1,Yano}.

An $2n$-dimensional smooth manifold $M$ is called to be {\it
almost complex} if the {\it almost complex structure} $J:TM \to
TM$, $J^2=-{\rm id}|_{TM}$ is defined in its tangent bundle. A
ten\-sor field $N$ of the type (1,2) on $M$ defined by the formula
     $$
N(X,Y) = 2 ([JX,JY]- [X,Y] - J[X,JY] - J[JX,Y]),
     $$
for any vector fields $X,Y$ is called {\it torsion} of 
$J$. If $N=0$ then $J$ is called to be {\it complex
structure}. In this case $(M,J)$ is called {\it
complex manifold}.

Let $(M,J)$ be a complex manifold. According to the 
Newlander-Nirenberg theorem \cite{kobn},
there exists an unique complex analytic manifold $M^c$
coinciding with $M$ as
topological space and such that its complex analytic structure induces
the complex structure $J$ and the structure of differential manifold
on $M$.

The tangent bundle $TM^c$ is {\bf C}-linear isomorphic to the bundle $TM$
with the structure of complex bundle induced
by $J$ so that there is a canonical {\bf C}-linear bundle isomorphism
  \bqq{1a}
TM \otimes_\rrr \cc \cong TM^c  \oplus \ol{TM^c}
  \eeq
where $TM\otimes_\rrr \cc$ is the complexification of $TM$
and the bar denotes the complex conjugation.

Let $(U, z^{\alpha})$, $\alpha=1,...,n$ be a chart on $M^c$.
If $M$ is the complex manifold
corresponding to $M^c$ then we shall say that $(U,z^\alpha,
\ol{z^\alpha})$, $\alpha=1,...,n$  (or simply $(U,z,\ol{z})$) is 
{\it complex chart} on $M$. 
Because of the isomorphism (\ref{1a}) vector fields $\partial_\alpha
\equiv \partial / \partial z^{\alpha}, \partial_{\ol \alpha} \equiv
\partial / \partial \overline{z^{\alpha}}$, $\alpha=1,...,n$ define a
basis in \tmc. Any real tensor field $T$ on $M$ can be uniquely
extended to the smooth field of elements of the "complexified" tensor algebra
     $$
\tilde{\bf T}_p M \equiv \bigoplus \limits^{\infty}_{k_i=1} ( (T_p
M^c)^{\otimes k_1}\otimes (\ol{T_p M^c})^{\otimes k_2} \otimes (T^*_p
M^c)^{\otimes k_3}\otimes (\ol{T^*_p M^c})^{\otimes k_4}).
     $$
In the coordinate basis $(\pl{\alpha}$, $\pl{\ol\alpha})$, $\alpha=1,...,n$
this extension has the form
     $$
T = T^{i_1 ... i_{r}}_{j_1 ... j_s} \partial_{i_1} \otimes ... \otimes
\partial_{i_{r}} \otimes dz^{j_1} \otimes ... \otimes dz^{j_s}, \quad
T^{\ol{i_1}...\ol{i_{r}}}_{\ol{j_1}... \ol{j_s}} = \ol{T^{i_1 ...
i_{r}}_{j_1 ... j_s}}.
     $$
Here the Latin indices varied from $1$ to $2n$ run over the sets of bared
$(\ol \alpha, \ol \beta, \ol \gamma,...)$ and unbarred $(\alpha,
\beta,\gamma,...)$ Greek indices varied from $1$ to $n$. 

In particular, the complex structure $J$ can be uniquely extended to
{\bf C}-linear endomorphism in \tmc. The action of complex structure
on the elements of coordinate basis 
is defined by the formulae
$J\partial_\alpha= i\partial_\alpha$, $J\partial_{\ol \alpha}=
-i\partial_{\ol \alpha}$.

Let us call {\it holomorphic transformation} a coordinate
transformation of the form $z'^{\alpha}=w^{\alpha}(z)$,
${z'}^{\ol\alpha}=\ol{w^\alpha (z)}$ where $w^{\alpha}(z)$ are complex
analytic functions. Let $X$ be a real vector field. If the Lie derivative
$L_X J$ is equal to zero then $X$ is called to be {\it holomorphic
vector field}. The condition $L_X J=0$ in a complex chart $(U,
z,\ol{z})$ yields $\partial_{\ol\nu}
\xi^\mu= \partial_{\nu} \xi^{\ol\mu} =0$, $\mu,\nu =1,...,n$. Using
the holomorphic coordinate transformations, in a vicinity of a
regular point any holomorphic vector field
can be reduced to the form $X=\pl 1 + \pl{\ol 1}$.

A complex manifold $(M,J)$ is called {\it
K\"ahler manifold} if a pseudo Riemannian metric $g$ can be defined on $M$
satisfying \cite{kobn,Yano}
     \bqq{sogl1}
g(JX,JY)=g(X,Y), \qquad \nabla_X J = 0
     \eeq
for any vector fields $X,Y$. Here $\nabla$ is the Levi-Civita connection of the metric $g$. The 2-form
    \bqq{333}
\Omega(X,Y)=g(JX,Y)
    \eeq
is called {\it fundamental} 2-{\it form} of K\"ahler manifold $M$.
From Eqs.~(\ref{sogl1}), (\ref{333}) and the condition $J^2 = -{\rm
id}|_{TM}$ it follows that $\Omega$ is closed: $d\Omega=0$.

Let $(U,z,\ol z)$ be a complex chart on $(M,g,J)$.
Then the components of the metric $g$, the complex structure $J$ and the fundamental
2-form $\Omega$ in the coordinate basis are defined by the conditions
   \bqq{mherm}
g_{\alpha\ol\beta}=\ol {g_{\ol\alpha\beta}},\qquad
g_{\alpha\beta}=g_{\ol\alpha\ol\beta}=0,
   \eeq
   \bqq{comJ}
J^\alpha_{\beta}=- J_{\ol\beta}^{\ol\alpha}=i\delta^\alpha_\beta,\qquad
J^\alpha_{\ol\beta} = J_{\beta}^{\ol\alpha} = 0,
   \eeq
   \bqq{com-omega}
\Omega_{\alpha\ol\beta}=\ol{\Omega_{\ol{\alpha} \beta}}= 
i g_{\alpha\ol\beta}, \qquad
\Omega_{\alpha\beta}=\Omega_{\ol\alpha\ol\beta}=0
\eeq
while the condition $d\Omega=0$ takes the form
   \bqq{zamkn}
\partial_\alpha g_{\beta\ol\gamma}=\partial_\beta g_{\alpha\ol\gamma},
\qquad
\pl{\ol\alpha} g_{\ol\beta\gamma} = \pl{\ol\beta}g_{\ol\alpha\gamma}.
   \eeq
From here it follows that in $U$ exists a real-valued function $\Phi$ obeying
   \bqq{kahl}
g_{\alpha\ol\beta}=\partial_{\alpha}\partial_{\ol\beta}\Phi.
   \eeq
This function is called {\it K\"ahler potential} of the metric $g$. It is
defined up to the {\it gauge transformations}
  \bqq{gau}
\Phi' = \Phi +f(z) +\ol{f(z)}.
  \eeq
where $f$ is an appropriate holomorphic function. 
From (\ref{mherm})--(\ref{kahl}) it follows that the only non-zero
Christoffel symbols and Riemann tensor of the metric $g$ are
\bqq{cris}
\Gamma^\alpha_{\beta\nu}= \ol{\Gamma^{\ol\alpha}_{\ol\beta
\ol\nu}}= g^{\alpha\ol\mu}\partial_\beta g_{\ol\mu\nu}
\eeq
while non-zero components of Ricci tensor $Ric$ are
defined by the conditions
    \bqq{riem}
R^{\alpha}_{\beta\mu\ol\nu} = \ol{ R^{\ol\alpha}_{\ol\beta\ol\mu\nu} }
= -R^{\alpha}_{\beta\ol\nu\mu} =
-\ol{R^{\ol\alpha}_{\ol\beta\nu\ol\mu} } = -\partial_{\ol\nu}
\Gamma^{\alpha}_{\beta\mu},
    \eeq
    \bqq{ric}
R_{\alpha\ol\beta} = \partial_{\alpha}\partial_{\ol\beta} \: \mbox{ln}
(\det (g_{\mu\ol\nu})), \qquad R_{\alpha\ol\beta} =
\ol{R_{\ol\alpha\beta}}.
    \eeq


\small
\section{$H$-projective mappings of K\"ahler manifolds}
\normalsize

A smooth curve $\gamma: t \mapsto x(t)$ on a K\"ahler
manifold $(M,g,J)$ of real dimension $2n>2$ is called to be {\it $H$-planar
curve} if its tangent vector $\chi \equiv dx/dt$ satisfies the equations
     $$
\nabla_{\chi} \chi = a(t)\chi +  b(t) J(\chi)
     $$
where $a(t)$ and $b(t)$ are functions of the parameter $t$.

Let us consider two K\"ahler manifolds $M$, $M'$ with metrics $g$,
$g'$ and complex structures $J$, $J'$. A diffeomorphism $f: M \to M'$
is called $H$-{\it projective mapping} if for any $H$-planar curve
$\gamma$ in $M$ the curve $f\circ \gamma$ is $H$-planar curve in $M'$.
If a pair of K\"ahler manifolds $M$ and $M'$ admit a
non-affine $H$-projective mapping
$f:M\to M'$ then we shall say that these two
manifolds are {\it $H$-projectively
equivalent.} Any non-affine $H$-pro\-jec\-ti\-ve mapping
preserve the complex
structure, i.e. $f_* \circ J = J' \circ f_* $ \cite{skm}.

Necessary and sufficient condition for a diffeomorphism
$f$ to be $H$-projective
mapping can be expressed by the equation
\cite{sin1,Yano}
   \bqq{sh-bis1}
f_{*}^{-1} (\nabla' {}_{f_* X} (f_* Y)) - \nabla_X Y=
p(Y)X + p(X)Y - p(JX) JY - p(JY) JX
   \eeq
where $p$ is a closed 1-form ($dp=0$) on $M$ and $\nabla$,
$\nabla'$ are the covariant derivatives with respect to Levi-Civita connections
of the metrics $g$, $g'$. If, in particular,
$p=0$, then $H$-projective
mapping preserves the connection and is affine.
We shall consider further only non-affine, i.e. proper
$H$-projective mappings. The condition (\ref{sh-bis1}) is equivalent to
the following equation
         $$
(\nabla \tg)(X,Y,Z) = 2p(Z)\tg (X,Y) + p(X)\tg (Y,Z) + p(Y)\tg (X,Z) -
         $$
         $$
p(JX)\tg (Y,JZ) - p(JY)\tg (X,JZ)
         $$
where $\tg= f_* g'$ and $X,Y,Z$ are vector fields on $M$.
In a complex coordinates, setting $Y=\partial_\alpha$,
$Z=\partial_{\ol\beta}$ and $W =
\partial_\gamma$, we get with the help of (\ref{mherm}) and (\ref{comJ})
         \bqq{hp2}
g'_{\alpha\ol\beta,\gamma} = 2 g'_{\alpha\ol\beta}\psi_{,\gamma} + 2
g'_{\gamma\ol\beta}\psi_{,\alpha}, \qquad g'_{\alpha\beta,\gamma} =
g'_{\alpha\beta,\ol\gamma}=0
         \eeq
where $g'_{ij}$ are components of the pullback $f^* g'$,
comma denotes the covariant derivation and $p=\psi_{,i} dx^i$. Note, that
$f^* g'$ is a K\"ahler metric on $(M,J)$ because
$f$ preserves the complex structure.
Hence, $g'_{ij}$ obey the conditions similar to (\ref{mherm}) and
(\ref{zamkn}).

Using the {\it Sinyukov's transformation} \cite{sin1}
         \bqq{sin}
a_{\alpha\ol\beta}= \ol{a_{\ol \alpha\beta}} =
e^{2\psi}g'^{\lambda\ol\mu}g_{\alpha\ol\mu} g_{\lambda\ol\beta},
\qquad a_{\alpha\beta}=a_{\ol\alpha\ol\beta}=0, \qquad
g^{\alpha\ol\beta}= e^{-2\psi } a^{\alpha\ol\beta}
         \eeq
where $a^{\alpha\ol\beta} = a_{\mu\ol\lambda}g^{\alpha\ol\lambda}
g^{\mu\ol\beta}$ and $(g'^{\alpha\ol\beta})= (g'_{\alpha\ol\beta})^{-1}$,
we can write (\ref{hp2}) in the form
         \bqq{hpa}
a_{\alpha\ol\beta,\gamma} = \lambda_\alpha g_{\gamma\ol\beta}, \qquad
a_{\alpha\ol\beta,\ol\gamma} = \lambda_{\ol\beta} g_{\gamma\ol\alpha}
         \eeq
where
         $$
\lambda_{\alpha} = \ol{\lambda_{\ol\alpha}} =
- 2 \psi_{,\nu} e^{2 \psi}g'^{\nu\ol\mu} g_{\alpha\ol\mu}.
         $$
Transvecting (\ref{hpa}) with $g^{\alpha\ol\beta}$, we find
         \bqq{lambda}
\lambda_\gamma =\ol{\lambda_{\ol\gamma}} = \frac{1}{2} \pl\gamma (g^{ij} a_{ij}) =
\partial_{\gamma} \lambda,
         \qquad
\lambda = a_{\alpha\ol\beta} g^{\alpha\ol\beta}.
         \eeq
From here it follows, that $\lambda_i dz^i=d\lambda$
for a real function $\lambda$.

The integrability conditions of (\ref{hpa}) follows from the Ricci
identity
         \bqq{ui-ij}
2a_{kl,[ij]} =a_{sl}R^s_{kij} + a_{ks} R^s_{lij}.
         \eeq
For $(ijkl)= (\gamma\ol\nu\alpha\ol\beta)$ and
$(\gamma\nu\alpha\ol\beta)$
using (\ref{riem}) we get
         \bqq{ui-1}
a_{\mu\ol\beta} R^{\mu}_{\alpha\gamma\ol\nu}+
a_{\alpha\ol\mu}R^{\ol\mu}_{\ol\beta\gamma\ol\nu}=
g_{\gamma\ol\beta}\lambda_{\alpha,\ol\nu}-
g_{\alpha\ol\nu}\lambda_{\ol\beta,\gamma},
         \eeq
         \bqq{ui-2}
g_{\gamma\ol\beta}\lambda_{\alpha,\nu}-
g_{\nu\ol\beta}\lambda_{\alpha,\gamma}=0.
         \eeq
The remaining integrability conditions hold identitically or can be
obtained from (\ref{ui-1}) and (\ref{ui-2}) by complex conjugation.
Contracting (\ref{ui-1}) with $g^{\alpha\ol\nu}$ we find
         $$
- a_{\mu\ol\beta} R^{\mu}_{\gamma}+
a_{\alpha\ol\mu}{R^{\ol\mu}_{\ol\beta\gamma}}^{\alpha}=
g_{\gamma\ol\beta} g^{\alpha\ol\nu} \lambda_{\alpha,\ol\nu}- n
\lambda_{\ol\beta,\gamma}.
         $$
From here, using (\ref{lambda}) and the identity $a_{\alpha\ol\mu}
{R^{\ol\mu}_{\ol\beta\gamma}}^{\alpha}= a_{\alpha\ol\mu}
{R^{\alpha}_{\gamma\ol\beta}}^{\ol\mu}$, it is easy to derive
         \bqq{zvez}
a_{\mu}^{\nu} R^{\mu}_{\gamma}- a_{\gamma}^{\mu} R^{\nu}_{\mu}=0.
         \eeq

Transvecting
(\ref{ui-2}) with $g^{\gamma\ol\beta}$ we find
$(n-1)\lambda_{\alpha,\nu} =0$ which means that $\lambda_{\alpha,\nu}
=0$ and $\lambda^{\alpha}_{,\ol\nu} =0$, or, because
$\Gamma^\alpha_{\ol\nu i} =0$,
    \bqq{hol-lam}
\partial_{\ol\nu} \lambda^{\alpha} =0, \qquad \partial_{\nu}
\lambda^{\ol\alpha} =0.
    \eeq
So, we come to the conclusion that $\Lambda = \lambda^i \partial_i$ is
a holomorphic vector field. Using the holomorphic coordinate transformations
$\Lambda$ can be reduced to the form
    \bqq{hol-form}
\Lambda = \pl1 +\pl{\ol 1}, \qquad \lambda^\alpha = \delta^\alpha_1,
\qquad \lambda^{\ol\alpha} = \delta^\alpha_1.
    \eeq

    \begin{The}  \label{sh-th1}
Let $f$ be a non-affine $H$-projective mapping of a K\"ahler manifold
$(M,g)$ on a K\"ahler manifold $(M',g')$. Let also $d\lambda =
\lambda_{\alpha} dz^\alpha+ \lambda_{\ol\alpha} dz^{\ol\alpha}$ be the
exact 1-form defined by Eqs.~{\rm (\ref{hp2}) -- (\ref{lambda})}. Then
the real vector field $J\Lambda = i\lambda^{\alpha} \pl\alpha -
i\lambda^{\ol\alpha} \pl{\ol\alpha}$ is infinitesimal isometry of
$M$, i.e. the Killing equations hold: $L_{J\Lambda} g =0$.
     \end{The}

\noindent {\bf Proof:} Using Eqs.~(\ref{cris}), (\ref{lambda}) and
(\ref{hol-lam}) we find
     $$
-i\lambda_{\ol\beta,\alpha} +i\lambda_{\alpha,\ol\beta} = -i
\pl{\alpha}\pl{\ol\beta} (a_{\nu\ol\mu} g^{\nu\ol\mu})+
 i \pl{\ol\beta}\pl{\alpha} (a_{\nu\ol\mu} g^{\nu\ol\mu})=0,
     $$
     $$
i\lambda_{\beta,\alpha} + i\lambda_{\alpha,\beta} = 0, \qquad
-i\lambda_{\ol\beta,\ol\alpha} - i\lambda_{\ol\alpha,\ol\beta} = 0,
     $$
or $L_S g_{ij} \equiv S_{i,j} + S_{j,i} =0$, where $S=J\Lambda$
and $S_i =g_{il} S^l$. Since $f$ is non-affine mapping the vector
field $\Lambda\ne 0$ and $S$ is the infinitesimal isometry. {\it Q.E.D.}

If we make $\Lambda=\pl{1} + \pl{\ol 1}$ (see (\ref{hol-form})),
then the Killing equations take the form
    \bqq{nondep1}
(\pl1 -\pl{\ol 1}) g_{\alpha\ol\beta} = 0.
    \eeq

    \begin{Lem} \label{sh-lm1}
If a K\"ahler manifold $(M,g,J)$ admits an infinitesimal isometry
$X$ which is a holomorphic vector field,
then the K\"ahler potential of $g$ can be reduced to the form
      \bqq{sh-bis2}
\Phi = \Phi (z^1 +z^{\ol 1}, z^2, z^{\ol 2}, ...).
      \eeq
      \end{Lem}

\noindent{\bf Proof:}
Any holomorphic vector field can be locally reduced to the form $X=i\dmin$.
Then the Killing equations take the form (\ref{nondep1}).
From here using (\ref{kahl}) we find
      \bqq{nondep2}
(\pl 1  -\pl{\ol  1})  g_{\alpha\ol\beta}  =  \pl{\alpha}\pl{\ol\beta}
\dmin\Phi =0.
      \eeq
Hence, $\dmin \Phi= f(z) + h(\ol z)$ where $f$ is a holomorphic
function and $h$ is an antiholomorphic function. Similarly, because $\Phi$
is real we have $(\pl 1 - \pl{\ol 1}) \Phi = -
\ol{(\pl 1 -\pl{\ol 1}) \Phi}$ and $h(\ol z) =- \ol{f(z)}$. Let us
change the K\"ahler potential using the gauge transformations of the form
      $$
\Phi = \Phi' +\int f(z) dz^1 + \ol{\int f(z) d z^1}= \int f(z)dz^1
-\int h(\ol z) d \ol{z^1}.
      $$
Substituting this expression in (\ref{nondep2}), we obtain $(\pl 1
-\pl{\ol 1}) \Phi' =0$. From here we find $\Phi'= \Phi'
(z^1 + z^{\ol 1}, z^2, z^{\ol 2}, ...)$. {\it Q.~E.~D.}

Let a K\"ahler manifold $(M,g)$ admits a non-affine $H$-projective
mapping. Then, according to Theorem~\ref{sh-th1} and
Lemma~\ref{sh-lm1} the K\"ahler potential can
be reduced to the form (\ref{sh-bis2}).
In this case (\ref{hpa}) yields
  \bqq{hpa-mod1}
a^{\alpha}_{\beta,\ol\gamma}= \pl{\ol\gamma} a^{\alpha}_{\beta} =
\delta^{\alpha}_{1} g_{\beta\ol\gamma}, \qquad a^{\alpha}_{\beta} =
g^{\alpha\ol\sigma} a_{\beta\ol\sigma},
  \eeq
  \bqq{hpa-mod2}
a^{\alpha}_{\beta,\gamma}= \lambda_{,\beta} \delta^\alpha_\gamma.
  \eeq
Integrating the first equation in (\ref{hpa-mod1}), we find
  \bqq{hpa-sol1}
a^\alpha_\beta = \delta^\alpha_1
\pl\beta \Phi + h^\alpha_\beta
      \eeq
where $h^\alpha_\beta$ are holomorphic functions.
From (\ref{lambda}) $\lambda =
\pl 1 \Phi + h^\alpha_\alpha$. Since $\lambda$ and
$\pl{1} \Phi$ are real we get
    \bqq{rho1}
\lambda = \pl 1 \Phi +n\rho, \qquad
h^\alpha_\alpha =h^{\ol\alpha}_{\ol\alpha} \equiv n\rho =
\mbox{const}
     \eeq
where we have used the fact that a holomorphic function is real iff it
is constant.
Substituting (\ref{rho1}) in (\ref{hpa-mod2}), we get
    \bqq{hpa-mod3}
a^{\alpha}_{\beta,\gamma}= g_{\beta \ol 1} \delta^\alpha_\gamma.
    \eeq

In the next section we shall consider this equation for the
case of a four-dimensional K\"ahler manifold.


\small
\section{Non-Einstein manifolds of dimension four}
\normalsize

Let $(M_{4},g,J)$ be a non-Ein\-stein ($Ric \neq\kappa g$)
K\"ah\-ler manifold of dimension ${\mbox{dim}}_\rrr M_{4} =4$.
Let $M_{4}$ admits a non-af\-fi\-ne $H$-pro\-jec\-ti\-ve
mapping on a K\"ahler manifold $(M'_{4},g',J)$ and let $a$ be the tensor
field defined by (\ref{sin}). We introduce tensor field
$b= L_{J\Lambda}a$ where $J \Lambda$ is the
infinitesimal isometry defined by Theorem~\ref{sh-th1}.

According to (\ref{hpa-sol1}) and Lemma \ref{sh-lm1},
in a complex coordinates where
   \bqq{lam-jlam}
\Lambda = \pl 1+\pl{\ol 1}, \qquad 
J \Lambda = i\dmin
   \eeq
we have 
   \bqq{hpa-sol2}
a^\alpha_\beta = \ol{a^{\ol\alpha}_{\ol\beta}} = \delta^\alpha_1
\pl\beta \Phi + f^\alpha_\beta (z^1,z^2) +\rho \delta^\alpha_\beta,
\qquad f^\alpha_\alpha= f^{\ol\alpha}_{\ol\alpha} = 0,
   \eeq
   \bqq{potl}
\Phi = \Phi(z^1 +z^{\ol 1},z^2,z^{\ol 2})
   \eeq
where $f^\alpha_\beta \equiv h^\alpha_\beta-\rho \delta^\alpha_\beta$
are holomorphic functions and $\Phi$ is the K\"ahler potential. From here
    \bqq{hpab-sol}
b^\alpha_\beta=\ol{b^{\ol\alpha}_{\ol\beta}}=i\dmin a^\alpha_\beta =
i\pl 1 f^\alpha_\beta, \quad b^{\ol\alpha}_\beta=
b^\alpha_{\ol\beta}=0, \quad
b^\sigma_\sigma=b^{\ol\sigma}_{\ol\sigma}=0,
    \eeq

Admissible coordinate and gauge transformations which don't change
the form of vector field $\Lambda=\pl 1 +\pl{\ol 1}$ and the form
(\ref{potl}) of K\"ahler potential are
      \bqq{free1}
{z'}^1 = z^1+l(z^2), \qquad {z'}^2 = m(z^2),
      \eeq
      \bqq{free-gau}
\Phi'= \Phi +r\cdot (z^1 + \ol{z^1}) + u(z^2) + \ol{u(z^2)}, \qquad
r\in \rr
      \eeq
where $l$, $m$ and $u$ are holomorphic functions depending on $z^2$ only.
Taking the Lie derivative along $J\Lambda$ from both parts of (\ref{zvez}),
we get
      \bqq{zvez-b}
b_{\mu}^{\nu} R^{\mu}_{\gamma}- b_{\gamma}^{\mu} R^{\nu}_{\mu}=0.
     \eeq

Using this formula it is possible to prove the following

\begin{Lem}\label{sh-lm2}
If a non-Einstein four-dimensional K\"ahler manifold $M_4$ admits a
non-affine $H$-projective mapping, then in a neighborhood of each
point $p \in M_4$ exist complex coordinates in which the following
relations hold
   \bqq{sh48}
a^\alpha_\beta =\delta_1^\alpha \pl\beta\Phi +f_\beta^\alpha(z^2)
+\rho \delta_\beta^\alpha, \qquad \dmin\Phi =0.
   \eeq
   \end{Lem}

We have placed the proof, which is rather long and technical, in
Appendix~A so as not interrupt exposition.

Admissible coordinate and gauge transformations not changing
(\ref{sh48}) are defined by the formulas (\ref{free1}) and
(\ref{free-gau}). Using these transformations one can reduce
$f^\alpha_\beta$ to one of the following forms:

\noindent
a) $f^\alpha_\beta
=\delta^\alpha_2\delta_\beta^1$ for $f^2_1 \neq 0$,

\noindent
b) $f^\alpha_\beta = \mu \ve\beta\delta^\alpha_\beta$, $\ve\beta
=(-1)^{\beta+1}$ for $f^2_1= 0$.

If we admit the first possibility then we come to the contradiction with
the assumption that $M_4$ is non-Einstein manifold (see proof in
Appendix~B).

In the second case we have
    $$
a^\alpha_\beta = \delta^\alpha_1 \pl\beta \Phi+
\mu \ve\beta \delta^\alpha_\beta + \rho \delta^\alpha_\beta,
\qquad \ve\beta = (-1)^{\beta+1},
\qquad
\mu=\mu (z^2)
    $$
and, from (\ref{zvez})
    \bqq{sh52}
R^2_1 = 0, \qquad (\pl 1\Phi +2\mu) R^1_2 + \pl 2 \Phi (R^2_2-R^1_1)=0.
    \eeq
Using the symmetry and the reality of $a$ we find
    \bqq{sh53}
g_{1\ol 1}(\mu -\ol\mu)=0,
    \eeq
    $$
g_{2\ol 1}\pl{\ol 2}\Phi - g_{1\ol 2}\pl 2 \Phi = g_{2\ol
2}(\ol\mu-\mu),
    $$
    \bqq{sh54}
g_{1\ol 1}\pl{\ol 2}\Phi - g_{1\ol 2}\pl 1 \Phi = g_{1\ol
2}(\ol\mu+\mu).
    \eeq
If $g_{1\ol 1}\ne 0$ then from (\ref{sh53}) it follows that $\mu=\ol\mu$.
In the case $g_{1\ol 1} =g_{2\ol 2}= 0$ we find from (\ref{sh54}) that
$\pl 1 \Phi =-(\mu+\ol\mu)$. Similarly,
by (\ref{zamkn}) and (\ref{kahl}), we get
     \bqq{sh55}
\pl 1 g_{1\ol 2}=\pl 1 g_{2\ol 1}=\pl 1 g_{2\ol 2}=0.
     \eeq
Therefore, from (\ref{ric}) we find $R^1_1=R^2_2=0$  and from (\ref{sh52})
it is easy to get 
$R^1_2 (\mu - \ol \mu)=0$. Hence $R^1_2=0$ for $\mu \neq \ol\mu$ and
$R^i_j=0$. So we find that $M_4$ is
an Einstein manifold that contradicts to
our initial assumption. Therefore, $\mu=\ol\mu=$const.

Making the transformation $\rho \to \rho-\mu$, $\Phi \to \Phi+2\mu(z^1
+\ol{z^1})$, we reduce $a^\alpha_\beta$ to the form
    \bqq{sh56}
a^\alpha_\beta = \delta_1^\alpha \pl\beta\Phi
+\rho\delta^\alpha_\beta.
    \eeq
The reality and the symmetry of the tensor $a$ imply
    \bqq{sh57}
\pl 2\Phi =\varphi\pl 1\Phi
    \eeq
where $\varphi=\varphi(z^2,\ol{z^2})$ is a complex function.
Using (\ref{kahl}) we get
    $$
g_{2\ol 1} = \varphi g_{1\ol 1},
\qquad g_{2\ol 2}= \pl{\ol 2}\varphi \pl 1\Phi+ \varphi\ol\varphi g_{1\ol 1}.
    $$
Because $g_{2\ol 2}$ is real, $\pl{\ol
2}\varphi\pl 1\Phi=\pl 2\ol\varphi\pl{\ol 1}\Phi$, and by
Lemma~\ref{sh-lm1}
    \bqq{sh58}
\pl{\ol 2}\varphi= \pl 2\ol\varphi.
    \eeq
This equation can be interpreted as the integrability
condition of the system
    \bqq{sysur}
\varphi = \pl 2 F, \qquad \ol\varphi = \pl{\ol 2} F
    \eeq
where $F$ is a real function depending only on $z^2$ and $z^{\ol 2}$. If
the equation (\ref{sh58}) holds, then (\ref{sysur}) has a solution
$F$. Substituting it in (\ref{sh57}), we find
    \bqq{sh59}
\pl 2\Phi= \pl 2 F\pl 1 \Phi
    \eeq
where $\pl 2\pl {\ol 2} F \neq 0$, because otherwise $\det \:
(g_{\alpha\ol\beta})= 0$.

Because of (\ref{sh56}), (\ref{sh57}) and (\ref{sh59}) the equation
(\ref{hpa}) holds identically. 
It means that any K\"ahler manifold whose K\"ahler
potential in any complex chart obeys
the equations (\ref{sh56}), (\ref{sh57}) and (\ref{sh59})
admits non-affine $H$-projective
mappings.

Now we find general solution of the equation (\ref{sh59})
for an appropriate function real $F(z^2, \ol{z^2})$.
Let $\tilde F (z^2,\ol{z^2})$ be a real function
functionally independent from $F$.
Rewriting (\ref{sh59}) in the variables
$u=F(z^2,\ol{z^2})$ and $v=\tilde F (z^2,\ol{z^2})$, we find
      \bqq{*1}
\pl u\Phi+\frac{\pl 2 \tilde F}{\pl 2 F}\pl v\Phi=\pl 1\Phi.
      \eeq
From here, taking into account the reality of the functions $F$, $u$
and $v$ as well as the identity $\dmin\Phi=0$, we find
      $$
(\frac{\pl 2\tilde F}{\pl 2 F}- \frac{\pl{\ol{2}}\tilde F}{\pl{\ol 2}
F})\pl{v}\Phi=0.
      $$
Since $F$, $\tilde F$ are functionally independent $\pl v\Phi=0$
and by (\ref{*1}) $\pl u \Phi - \pl 1 \Phi=0$.
Therefore, the general solution of (\ref{sh59})
has the form $\Phi ={\cal W}(z^1+\ol{z^1} +F(z^2, \ol{z^2}))$
where ${\cal W}$ is an appropriate real function of one real variable.

From these relations the main result now follows
      \begin{The} \label{sh-th2}
Let $f$ be a non-affine $H$-projective mapping of a non-Einstein
fo\-ur-di\-men\-si\-o\-nal K\"ah\-ler manifold $(M_4,g,J)$
on a K\"ahler manifold
$(M'_4,g',J)$. Then in a neighborhood of each point $p\in M_4$
exist complex coordinates $(z^{\alpha},z^{\ol\alpha})$,
$\alpha=1,\ldots,n$ in which K\"ahler potential $\Phi$ can
be chosen in the form
     \bqq{sh60}
\Phi ={\cal W} (z^1+\ol{z^1}+F(z^2,\ol{z^2})), \quad F=\ol F, \quad \pl 2
\pl{\ol 2} F\neq 0,\quad {\cal W}\neq \mbox{const}
     \eeq
and the components of the metric $g$ are defined by the formula
     \bqq{kahl-a}
g_{\alpha\ol\beta}=\partial_{\alpha}\partial_{\ol\beta}\Phi.
     \eeq
In the same coordinate system
the pullback $f^* g'$ of the metric $g'$ is defined by Eq.~{\rm
(\ref{sin})} where
     \bqq{sh69}
a_{\alpha\ol\beta}= \ol{a_{\ol\alpha\beta}} = \pl\alpha\Phi\pl
1\pl{\ol\beta}\Phi+ \rho\pl\alpha\pl{\ol\beta} \Phi,\quad
a_{\alpha\beta}=a_{\ol\alpha\ol\beta}=0, \quad \rho\in\rr.
     \eeq
     \end{The}


\small
\section{Generalized equidistant K\"ahler manifolds
\newline
and gravitational instantons}
\normalsize

A (pseudo)Riemannian manifold $(M,g)$ is called
{\it equidistant} \cite{sin1}
if it admits  a covector field $\varphi$ obeying the condition 
$(\nabla \varphi ) (X,Y)=\rho g(X,Y)$ where
$\rho$ is a smooth function and
$X,Y$ are appropriate vector fields on $M$.
If in (\ref{sh60}) ${\cal W}(x) =\exp (x)$ then (\ref{kahl-a}) defines
the metrics of an equidistant K\"ahler manifolds.
Conversely, it can be shown 
that the K\"ahler potential of any
equidistant manifold can be reduced to the form \cite{mik-ekv,skm}
   $$
\Phi(z^1,\ol{z^1},\ldots, z^n, \ol{z^n})=
\exp (z^1+\ol{z^1}+F(z^2,\ol{z^2},\ldots,z^n,\ol{z^n}))
   $$
for a real function $F$.

We now define a more general
class of K\"ahler manifolds then those of equidistant manifolds.
A K\"ahler manifold $M$ is called to be {\it generalized equidistant} if in
local complex coordinates its K\"ahler potential can be reduced to the form
    $$
\Phi ={\cal W} (z^1+\ol{z^1}+F(z^2,\ol{z^2},\ldots,z^n,\ol{z^n})),
\qquad F=\ol F.
    $$

Let us consider a four-dimensional generalized equdistant K\"ahler manifold
with the metric $g$ given by (\ref{sh60}) and the tensor field $a$ defined
by the equation (\ref{sh69}). As it was shown in the
previous section, Eq.~(\ref{hpa}) where $\lambda_\alpha= g_{\alpha \ol 1}$,
$\lambda_{\ol\alpha} = g_{\ol\alpha 1}$ holds identically for such
$g$ and $a$.
Therefore, we have the following

\begin{The} \label{gen-eqv-th}
Any four-dimensional generalized equidistant K\"ah\-ler
ma\-ni\-fold ad\-mits
a non-affine $H$-projective mapping. 
\end{The}

J.~Mike\v s
\cite{mik-ekv} have proved that equidistant K\"ahler manifolds admit
non-affine $H$-projective mappings and Theorem \ref{gen-eqv-th} confirms
this result for the case of four-dimensional manifolds. 

It is well-known that K\"ahler manifolds of constant holomorphic sectional
curvature admits $H$-projective mappings. 
It is easy to show that such manifolods are generalized equidistant with
        $$
\Phi = \ln (1 + \epsilon \exp (z^1+\ol{z^1}+ \ln (1+ \sum_2^n z^\alpha
\ol{z^\alpha}))), \qquad \epsilon = \pm 1
        $$
for non-zero holomorphic sectional curvature and
        $$
\Phi = \exp (z^1+\ol{z^1}+ \ln (1+ \sum_2^n z^\alpha
\ol{z^\alpha}))
        $$
in the flat case. In particular, $\cc\pp{}^n$ and $\cc^n$
are generalized equidistant manifolds.

It is possible also to
construct the following class of the generalized equidistant manifolds.
Let $N$ be an algebraic submanifold in $\cc^{n+1}$ defined by the equation
        $$
{\cal F}_N (z^2,\ldots,z^{n+1})=0
        $$
where the function ${\cal F}_N$ is a polynomial
invariant with respect to the action of the group
$\cc^*=\cc\backslash \{0 \}$ on $\cc^{n+1}$ by multiplications.
Then $M=N/\cc^*$ is a $n-1$-dimensional algebraic submanifold in
$\cc\pp{}^n$.
Taking in $\cc\pp{}^n$ K\"ahler metric with the potential
defined by the formula \cite{kobn}
        $$
\Phi=\ln (z^1 \ol{z^1}+z^2 \ol{z^2} + \ldots +z^{n+1} \ol{z^{n+1}})
        $$
it is easy to see that $M$ with induced metric is generalized
equidistant manifold.

We now consider the Einstein generalized equidistant manifolds
($Ric=\kappa g$).
In the case $\kappa=0$ the manifolds are Ricci-flat. Hence, they possess
hyper K\"ahler structure \cite{bess}. For any value of $\kappa$ the
Einstein-K\"ahler manifolds
have various important applications in theoretical
and mathematical physics
\cite{azk,bess,perry}. In particular, such manifolds describe field 
configurations of gravitational instantons \cite{perry}.
From the point of view of differential geometry the problem
of finding four-dimensional Einstein-K\"ahler manifolds is
also of great interest and leads to investigation of complex
Monge-Amp\`ere equation \cite{bess,Yau}.

Ein\-stei\-n-K\"ahler generalized
equidistant manifolds are distinguished by the condition
     \bqq{*2}
\exp (-\kappa \Phi)\pl 1 (\pl 1 \Phi)^2 \pl 2\pl {\ol 2} F = f(z) \ol{f(z)}
     \eeq
where $f(z)$ is an appropriate holomorphic function.
By the use of coordinate
transformations one can make $f(z)\ol{f(z)}=\mbox{const}$ or 
$f(z)\ol{f(z)}=\mbox{const} \exp (z^1+\ol{z^1})$.
For simplicity we restrict our further consideration
only to the first case.
In the first case we have
     $$
\exp (-\kappa {\cal W}) {\cal W}' {\cal W}'' \pl{2\ol 2}
F= \mbox{const} \ne 0.
     $$
Because $F$ depends on $z^2$, $\ol{z^2}$ only this equation
can be rewritten as
     \bqq{eq-W}
{\cal W}' {\cal W}'' \exp (-\kappa {\cal W})=\mbox{const},
\qquad \pl{2\ol 2} F=\mbox{const}\ne 0
     \eeq
whence
     \bqq{z2}
F(z^2,\ol{z^2})=\gamma z^2 \ol{z^2}+\tau (z^2+\ol{z^2})+\sigma
     \eeq
where $\gamma$, $\tau$ and $\sigma$ are real constants. 

For $\kappa=0$ (Ricci-flat case) after integration
of (\ref{eq-W}), we find
     \bqq{eq-W2}
{\cal W}=A(x+B)^{3/2}+C, \qquad
x=z^1+\ol{z^1}+F(z^2,\ol{z^2})
     \eeq
where $A$, $B$ and $C$ are some real constants.
After substituting (\ref{z2})
in (\ref{eq-W2}) and making the admissible coordinate change $z^1\to
z^1+(\tau^2-\sigma)/2$, $z^2\to z^2-\tau$, we obtain the following general
expression for K\"ahler potential $\Phi$
     $$
\Phi=A(z^1+\ol{z^1}+\gamma z^2\ol{z^2})^{3/2}
     $$
where the constant $C$ is omitted because it corresponds to the gauge
tran\-s\-for\-ma\-t\-i\-ons. From here it is easy to
get the following expression for the metric in complex coordinates
     $$
ds^2=\frac{3}{4}A (z^1+\ol{z^1}+\gamma z^2\ol{z^2})^{-1/2}
[dz^1 d\ol{z^1} + \gamma z^2 dz^1 d\ol{z^2} +
\gamma\ol{z^2}dz^2 d\ol{z^1} +
 $$
 \bqq{met-comp}
2\gamma(z^1+\ol{z^1}+\frac{3\gamma}{2} z^2\ol{z^2})\: dz^2 d\ol{z^2}].
 \eeq
Introducing the real coordinates
 $$
x=\frac{z^1+\ol{z^1}}{2}, \quad y=\frac{z^1 - \ol{z^1}}{2i},
\quad 
u=\frac{z^2+\ol{z^2}}{2}, \quad v=\frac{z^2-\ol{z^2}}{2i},
 $$
we find from (\ref{met-comp}) the following form of
the metric of four-dimensional Ricci-flat
generalized equidistant manifolds
 $$
ds^2=\frac{3}{4}A (x+\gamma (u^2+ v^2))^{-1/2}
[dx^2 +dy^2 +
 $$
 \bqq{met-real}
 2\gamma (u\; dx\: du + u\; dy\: dv -v\; dx\: dv +v\; dy\: du) 
 \eeq
 $$
2\gamma(x+\frac{3\gamma}{2} (u^2+v^2))(du^2 +dv^2)].
 $$
For the case $\kappa \ne 0$ we have from (\ref{eq-W}) and (\ref{z2})
  $$
{\cal W}' {\cal W}'' \exp (-\kappa {\cal W})=\mbox{const}.
  $$
After first integration of this equation we get
  \bqq{eqW1}
{\cal W}'=-\frac{1}{ \kappa } \: (B-Ae^{\kappa {\cal W} })^{1/3}
  \eeq
where $A$ and $B$ are some constants.
From here it is easy to find the metric coefficients
  \bqq{g11}
g_{1 \ol{1}}=\frac{-A}{3 \kappa}\: e^{\kappa {\cal W}} (B-A\: 
e^{\kappa {\cal W}} ){}^{-1/3},
  \eeq
  \bqq{g12}
g_{1 \ol{2}}=\frac{-A \gamma z^2}{3 \kappa} \:
e^{\kappa {\cal W}} (B-A \: e^{\kappa {\cal W}} ){}^{-1/3},
  \eeq
  \bqq{g22}
g_{2 \ol{2}} = \frac{-A \gamma z^2 \ol{z^2}}{3 \kappa}\:
e^{\kappa {\cal W}}\:
(B-A\: e^{\kappa {\cal W}} )^{-1/3} -
\frac{\gamma}{\kappa}\: (B-A\: e^{\kappa {\cal W}} ){}^{1/3}
  \eeq
where the function ${\cal W}$ has to be found from (\ref{eqW1}).
Integrating (\ref{eqW1}) in the case $B \ne 0$ we get
the following relation between the function ${\cal W}$
and its argument $x=z^1+\ol{z^1} + F(z^2,\ol{z^2})$
(here $F$ is given by (\ref{z2}))
  $$
x+C=\frac{-3}{ \kappa} \: (\: 
\frac{ \arctan (\frac{B^{1/3}+2 T}{\sqrt{3} B^{1/3} }) }{
\sqrt{3} B^{1/3} }-
  $$
  \bqq{Bnz}
\frac{\ln (-B^{1/3}+T)}{3 B^{1/3}} +
\frac{\ln ( B^{2/3}+B^{1/3} T+T^2)}{6 B^{1/3}})
  \eeq
where $T=(B-A\: e^{\kappa {\cal W}})^{1/3}$.
In the case $B=0$ from (\ref{eqW1}) it is easy to find
  \bqq{Bz}
{\cal W}=\frac{3}{\kappa}\: \ln (x+C)
  \eeq
where the additive constant $\frac{3}{\kappa}$ $\ln
(\frac{A^{1/3}}{3})$ is not written.

The equations (\ref{met-real}), (\ref{g11})--(\ref{Bz}) define
the metrics of Einstein generalized equidistant manifolds. The manifolds
of this type can 
be interpreted as field configurations of gravitational instantons.

\setcounter{section}{0}
\renewcommand{\theequation}{\Alph{section}.\arabic{equation}}

\small
\appsection
\normalsize

Here we provide the proof of Lemma~\ref{sh-lm2}.

It follows from (\ref{hpab-sol}) that $b^\alpha_\beta$ depend only on
$z$. Holomorphic coordinate transformations
don't change this result and can be used
to make $b^1_2=0$.

Let $b^1_2=0$, consider the following three possibilities in
Eq.~(\ref{zvez-b}).

\noindent
1) Let $b^2_1=0$ and the tensor field $b$ does not vanishes. Then either
$b^1_1=b^2_2=0$ that contradicts with the assumption about not
vanishing of tensor $b^i_j$ or, because $M_4$ is non-Einstein
manifold, $R^1_2=R^2_1=0$ and $R^1_1 \neq R^2_2$. In the last case it
is possible to find such functions $v_1$ and
$v_2$ that
   \bqq{sh39-bis}
b^\alpha_\beta = v_1 R^\alpha_\beta + v_2 \delta_\beta^\alpha.
   \eeq

\noindent
2) If $b^2_1 \neq 0$ and $b^1_1 \neq 0$, then from (\ref{zvez-b}) we have
   $$
R^1_2=0,\qquad \frac{R^1_1 -R^2_2}{2 b^1_1}= \frac{R^2_1}{b^2_1},
   $$
hence, $b^1_1-b^2_2=v_1(R^1_1-R^2_2)$, $b^2_1 = v_1 R^2_1=0$, $b^1_2=
v_1 R^1_2$ for some function $v_1$. By putting $b^\alpha_\beta = v_1
R^\alpha_\beta + \tilde b_\beta^\alpha $, we find $\tilde b^1_1 -
\tilde b^2_2 = \tilde b^1_2= \tilde b^2_1 =0 $ or $\tilde
b^\alpha_\beta = v_2\delta^\alpha_\beta$ where $v_2$ is some function
in $U$.

\noindent
3) At last, in the case $b^2_1\neq 0$, $b^1_1=b^2_2=0$ we get
$R^1_2=R^1_1-R^2_2=0$. Hence, it is possible to find 
functions $v_1$, $v_2$ such that that (\ref{sh39-bis}) holds. 
We come to the
conclusion that (\ref{sh39-bis}) describes all possible
cases. In the similar way the relations
      \bqq{sh39-1}
b^{\ol\alpha}_{\ol\beta} = \ol{v_1} R^{\ol\alpha}_{\ol\beta} +
\ol{v_2} \delta_\beta^\alpha, \qquad b^{\ol\alpha}_\beta = v_1
R^{\ol\alpha}_\beta + v_2 \delta_\beta^{\ol\alpha} \equiv 0.
      \eeq
can be obtained. From here because of the reality and the symmetry of
$a$, $b$ and $Ric$ it follows that $v_1$ and $v_2$ are
real-valued functions, i.e. (\ref{sh39-bis}), (\ref{sh39-1}) can be
rewritten in the form of one tensor relation
      $$
b^i_j = v_1 R^i_j + v_2 \delta_j^i.
      $$
Since $b^i_i=0$ we get $v_2 = - (v_1 R)/2n$, whence
    \bqq{sh40}
b^\alpha_\beta = v_1 (R^\alpha_\beta - \frac{R}{2n}
\delta_\beta^\alpha).
    \eeq
Because of (\ref{cris}), (\ref{hpa-mod3}) and (\ref{hpab-sol}) we have
$b^\alpha_{\beta,j} = 0$. Differentiating (\ref{sh40}) and 
denoting ${\cal A}=\ln v_1$, one can find
    $$
{\cal A}_{,j}=-\frac{(R^\alpha_\beta-\delta^\alpha_\beta R/2n)_{,j}}
{R^\alpha_\beta - \delta^\alpha_\beta R/2n}.
    $$
The right hand side of this relation doesn't depend on the variable
$y^1=\frac{1}{\sqrt{2}}(z^1 -z^{\ol 1})$, hence, its left hand side shouldn't 
depend too. Because ${\cal A}$ is real we have
    $$
{\cal A} =\tilde f (z^1 +z^{\ol 1},z^2,z^{\ol 2}) +i\tilde \tau\cdot
(z^1-z^{\ol 1}), \qquad \tilde \tau \in\rr,
    $$
    $$
v_1 = \exp {\cal A} = f (z^1 +z^{\ol 1},z^2,z^{\ol 2})\exp (i
\tau\cdot (z^1 - z^{\ol 1})),
   \qquad
\tau \in\rr.
   $$
Then from (\ref{sh40}) we get $b^\alpha_\beta = \exp (2i\tau z^1)
\tilde c^\alpha_\beta(z^2)$. Taking into account (\ref{hpab-sol})
   \bqq{sh41}
f^\alpha_\beta = -i\int b^\alpha_\beta dz^1 = \exp (2i\tau z^1)
c^\alpha_\beta(z^2) +d^\alpha_\beta(z^2), 
\qquad c^\alpha_\alpha = d^\alpha_\alpha =0.
    \eeq
Using this equation we find from (\ref{zvez}), (\ref{hpa-sol2})
    \bqq{sh43}
\delta^\alpha_1 \pl\mu \Phi R^\mu_\beta -\pl\beta\Phi R^\alpha_1
+d^\alpha_\mu R^\mu_\beta - d^\mu_\beta R^\alpha_\mu = 0,
    \eeq
    \bqq{sh44}
c^\alpha_\mu R^\mu_\beta - c^\mu_\beta R^\alpha_\mu = 0.
    \eeq

Let us consider the cases $c^2_1 \neq 0$ and
$c^2_1 =0$. Using the admissible transformations
(\ref{free1}) in the first case one can reduce
$c^\alpha_\beta$ to the form
   $$
c^{\alpha}_{\beta} = \delta^{\alpha}_1 \delta^2_{\beta} \phi+
\delta^{\alpha}_2 \delta^1_{\beta}
   $$
where $\phi$ is a holomorphic function depending on $z^2$ only.
Substituting this expression in (\ref{sh44}), we
find $\phi R^{2}_{1} =R^{1}_{2}$, $R^{1}_{1} =R^{2}_{2}$, whence, by
(\ref{sh43}) we have
   $$
(\pl{1} \Phi +2d^{1}_{1}) R^{1}_{2} =0, \qquad (\pl{1} \Phi
+2d^{1}_{1}) R^{2}_{1} =0.
   $$
If $R^1_2 \neq 0$ or $R^2_1 \neq 0$, then $\pl{1} \Phi = -2d^1_1$ is a
holomorphic function and $g_{1\ol 1}=g_{1\ol 2}=0$, hence, the metric
is degenerate. Therefore, $R^1_2 = R^2_1 = R^1_1 - R^2_2 = 0$ that
contradicts with our assumption that $M_4$ is non-Einstein manifold.

So we have $c^2_1=0$. In this case by the use of the admissible
coordinate transformations one can make
   \bqq{sh46}
c^\alpha_\beta = \ol{c^{\ol\alpha}_{\ol\beta}} = c^1_1
\delta_\beta^\alpha \ve\beta, \qquad \ve\beta =(-1)^{\beta+1},
   \eeq
   $$
d^\alpha_\beta = \ol{d^{\ol\alpha}_{\ol\beta}} = d^1_1
\delta_\beta^\alpha \ve\beta + \gamma (z^2) \delta^\alpha_1
\delta^2_\beta + \zeta \delta^\alpha_2 \delta^1_\beta, \quad \zeta=0,1.
   $$
After the gauge
transformation $\Phi\to \Phi +\int \gamma(z^2) dz^2 + \ol{\int
\gamma(z^2) d z^2}$ we find taking into account (\ref{hpa-sol2})
and (\ref{sh41})
   \bqq{sh47}
d^\alpha_\beta = \ol{d^{\ol\alpha}_{\ol\beta}} = d^1_1
\delta_\beta^\alpha\ve\beta+\zeta\delta^\alpha_2 \delta^1_\beta.
   \eeq
Substituting (\ref{sh46}) into (\ref{sh44}), we get $c^1_1 R^1_2 =0$,
$c^1_1 R^2_1 =0$, whence, $c^1_1=0$ or $R^1_2 = R_1^2 =0$. In the last
case from (\ref{sh43}) and (\ref{sh47}) follows $\pl 2 \Phi
(R^2_2-R_1^1)=0$. Since $\pl 2\Phi \neq 0$, we find $R^2_2-R_1^1=0$.
So $M_4$ is an Einstein manifold again, therefore, $c_1^1 = 0$ and
$f^\alpha_\beta = d^\alpha_\beta(z^2)$. Substituting this result
into (\ref{hpa-sol2}) prove the Lemma~\ref{sh-lm2}.


\small
\appsection
\normalsize

Here we prove that the condition $f^\alpha_\beta =\delta^\alpha_2
\delta^1_\beta$ contradicts with the assumption that the considered
K\"ahler manifold is non-Einstein.

From (\ref{zvez}) and (\ref{sh48}) we find
    \bqq{949}
R^1_2=\pl 2 \Phi R^2_1,  \quad R^1_2 \pl 1 \Phi + \pl 2 \Phi (R^2_2 -
R^1_1)=0.
    \eeq
Writing down the symmetry conditions of $a$ we obtain with the help of
(\ref{sh48}) the next formulae
    $$
g_{2\ol 1} =g_{1\ol 2}, \qquad g_{1\ol 1}\pl 2 \Phi = g_{2\ol
1}\pl{\ol 1}\Phi + g_{2\ol 2},
    $$
    $$
g_{1\ol 2} \pl 1 \Phi + g_{2\ol 2} = g_{1\ol 1} \pl 2 \Phi,
    $$
    $$
g_{1\ol 2}\pl 2 \Phi - g_{2\ol 1}\pl{\ol 2}\Phi \equiv g_{1\ol 2}(\pl
2 \Phi - \pl{\ol 2}\Phi) =0.
    $$
From the last equation it follows that either $\pl 2 \Phi =\pl{\ol 2}
\Phi$ or $g_{1\ol 2} = g_{2\ol 1} = 0$.

Let us first take $g_{2\ol 1} =g_{1\ol 2}= 0$, then from (\ref{zamkn})
the equality $\pl 1 g_{2\ol 2} = \pl 2 g_{1\ol 1} = 0$ follows, hence
$\pl 1\pl{\ol 2}\det (g_{\alpha\ol\beta})=0$ and, because of
(\ref{ric}) we find $R_{1\ol 2}=R_{2\ol 1}=0$, therefore,
$R_1^2=R^1_2=0$. Since $\pl 2\Phi \neq 0$,
from (\ref{949}) we get $R^1_1-R_2^2=0$, which means that
$M_4$ is Einstein manifold. We came to contradiction with our initial
assumption. Hence, in addition to the formula $\pl 1\Phi = \pl{\ol
1}\Phi$ we have $\pl 2\Phi = \pl{\ol 2}\Phi$. From here using
Eqs.~(\ref{kahl}) -- (\ref{ric}), it is possible to deduce that all
components of the metric tensor, Christoffel symbols and curvature
tensor are real. Then (\ref{zvez}) can be written as
    $$
R_{\alpha\ol\sigma}a^{\ol\sigma}_{\ol\beta} -
R_{\sigma\ol\beta}a^\sigma_\alpha = 0.
    $$
From here, putting $\alpha,\beta=1,2$ and using the identities $\pl 2
\Phi\neq 0$, $a^\alpha_\beta=a^{\ol\alpha}_{\ol\beta}$ and
$R_{\alpha\ol\beta}=R_{\ol\alpha\beta}$, we find
$R_{\alpha\ol\beta}=0$, hence, $Ric=0$ that contradicts
with the assumption that $M_4$ is non-Einstein. {\it Q.E.D.}

\end{document}